\documentclass[prl,twocolumn]{revtex4}
\usepackage{graphicx}
\usepackage{dcolumn}
\usepackage{bm}
\usepackage{subfigure}
\usepackage{amssymb}
\usepackage{array}
\usepackage{amsthm}
\usepackage{amsmath}
\usepackage{epsfig}
\usepackage{graphicx}
\usepackage{wasysym}
\usepackage{bbm}

\def \be{\begin{equation*}}
\def \ee{\end{equation*}}

\begin{document}

\title{Electronic Structure of Pyrochlore Iridates: From Topological Dirac
Metal to Mott Insulator}
\author{Xiangang Wan$^{1}$, Ari Turner$^{2}$, Ashvin Vishwanath$^{2,3}$,
Sergey Y. Savrasov$^{1,4}$ }
\affiliation{$^{1}$National Laboratory of Solid State Microstructures and Department of
Physics, Nanjing University, Nanjing 210093, China\\
$^{2}$Department of Physics, University of California, Berkeley, CA 94720\\
$^3$ Materials Sciences Division, Lawrence Berkeley National Laboratory, Berkeley CA 94720.\\
$^{4}$Department of Physics, University of California, Davis, One Shields
Avenue, Davis, CA 95616.}

\begin{abstract}
In 5\textit{d} transition metal oxides such as the iridates, novel
properties arise from the interplay of electron correlations and spin-orbit
interactions. We investigate the electronic structure of the pyrochlore
iridates, (such as Y$_{2}$Ir$_{2}$O$_{7}$) using density functional theory,
LDA+U method, and effective low energy models. A remarkably rich phase
diagram emerges on tuning the correlation strength $U$. The Ir magnetic
moment are always found to be non-collinearly ordered. However, the ground
state changes from a magnetic metal at weak $U$, to a Mott insulator at
large $U$. Most interestingly, the
intermediate $U$ regime is found to be a Dirac semi-metal, with vanishing
density of states at the Fermi energy. It also exhibits topological
properties - manifested by special surface states in the form of Fermi arcs,
that connect the bulk Dirac points. This Dirac phase, a three dimensional
analog of graphene, is proposed as the ground state of Y$_{2}$Ir$_{2}$O$_{7}$
and related compounds. A narrow window of magnetic `axion' insulator, with axion parameter $\theta=\pi$, may also be present at intermediate $U$. An
applied magnetic field induces ferromagnetic order and a metallic ground
state.
\end{abstract}

\maketitle




Previously, some of the most striking phenomena in solids, such as high
temperature superconductivity\cite{Orenstein} and colossal magnetoresistance%
\cite{CMR} were found in transition metal systems involving 3\textit{d}
orbitals, with strong electron correlations. Now it has been realized that
in 4\textit{d} and the 5\textit{d} systems, whose orbitals are spatially
more extended, a regime of intermediate correlation appears. Moreover, they
display significant spin-orbit coupling, which modifies their electronic
structure as recently verified in Sr$_2$IrO$_4$ \cite{Sr2IrO4-1}. This is a
largely unexplored domain, but already tantalizing new phenomena have been
glimpsed. For example, in the 5\textit{d} magnetic insulator, Na$_2$IrO$_3$%
\cite{Spin-liquid Na4I3O8 -exp}, a disordered ground state persists down to
the lowest measured temperatures, making it a prime candidate for a quantum
spin liquid\cite{S-L-3}.

It is known that strong spin-orbit interactions can lead to a novel phase of
matter, the topological insulator\cite{TI}. However, the experimental
candidates uncovered so far have weak electron correlations. Recently, it
was realized that the iridates are promising candidates to realize
topological insulators\cite{Nagaosa}, and that Iridium based pyrochlores in
particular \cite{Leon}, provide a unique opportunity to study the interplay
of Coulomb interactions, spin-orbit coupling and the band topology of solids.

The pyrochlore iridates, with general formula $A_{2}$Ir$_{2}$O$_{7}$, where $%
A=$ Yttrium, or a Lanthanide element, will be the main focus of this work.
Both the $A$ and Ir atoms are located on a network of corner sharing
tetrahedra \cite{pyrochlore review,Magnetic Pyrochlore Oxides}. Pioneering
experiments\cite{exp 2001 Ir-227} on the pyrochlore iridates, revealed an
evolution of ground state properties with increasing radius of the $A$ ion,
which is believed to tune electron correlations. While $A=$Pr is metallic, $%
A=$Y is an insulator as low temperatures. Subsequently, it was shown that
the insulating ground states evolve from a high temperature metallic phase,
via a magnetic transition\cite{Taira 2001 Ir-227,Matsuhira Ir-227}. The
magnetism was shown to arise from the Ir sites, since it also occurs in $A=$%
Y, Lu, where the $A$ sites are non-magnetic. While its precise nature
remains unknown, the absence of a net moment rules out ferromagnetism.

We show that electronic structure calculations can naturally account for
this evolution and points to a novel ground state whose properties are
described here. First, we find that magnetic moments order on the Ir sites
in a non-colinear pattern with moment on a tetrahedron pointing all-in or
all-out from the center. This structure retains inversion symmetry, a fact
that greatly aids the electronic structure analysis. While the magnetic
pattern remains fixed, the electronic properties evolve with correlation
strength. For weak correlations, or in the absence of magnetic order, a
metal is obtained, in contrast to the interesting topological insulator
scenario of Ref. \cite{Leon}. With strong correlations we find a Mott
insulator, with all-in/all-out magnetic order. However, for the case of
intermediate correlations, relevant to Y$_{2}$Ir$_{2}$O$_{7}$, the
electronic ground state is found to be an unusual \emph{Dirac semi-metal},
with linearly dispersing Dirac nodes \emph{at} the chemical potential.
Indeed, this dispersion is analogous to graphene\cite{Geim}, but occurs
inside a three dimensional magnetic solid. The small density of states leads
to a vanishing conductivity at low temperatures. The Dirac fermions here are
rather different from those in three dimensional semi-metals such as
elemental Bismuth, which are inversion symmetric and non-magnetic. Here, the Dirac fermions at a particular
momentum are described by a handedness (which is left or right handed), and
a two component wavefunction. They cannot be gapped unless they mix with a
fermion of opposite handedness. In contrast, Dirac fermions in Bismuth have
four component wavefunctions, no particular handedness, and are typically gapped.
Such a three dimensional electronic structure has, to our knowledge, not
been discussed before.

A key property of this Dirac semi-metal phase of two component Dirac
fermions, is unusual band topology, reminiscent of topological insulators.
Since the bulk fermi surface only consists of a set of momentum points,
surface states can be defined for nearly every surface momentum, and are
always found to occur on certain surfaces. They take the shape of `Fermi
arcs' in the surface Brillouin zone, that stretch between Dirac points.
Hence we term this phase \emph{topological Dirac metal}.

\begin{figure}[tbp]
\includegraphics[width=0.45\textwidth]{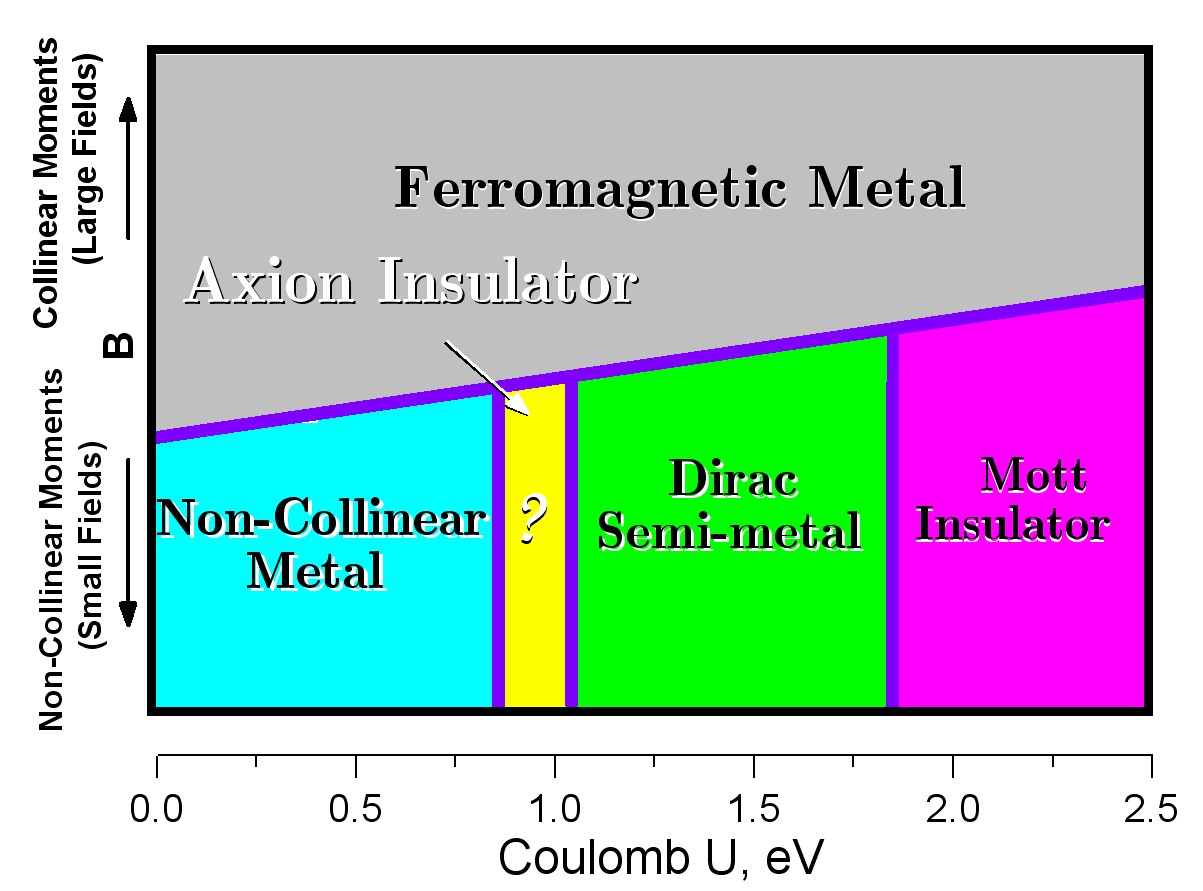}
\caption{Sketch of the phase diagram for pyrochlore iridiates
from our microscopic electronic structure calculation: Horizontal axis
corresponds to the increasing interaction among Ir 5d electrons (the scale
is obtained using LSDA+SO+U method) while the vertical axis corresponds
to external magnetic field which aligns the moments and triggers a
transition out of the zero field non-collinear "all-in/all-out" ground state.
We find normal magnetic metal, Mott insulator and Dirac semi-metal phases, and
also possibly an exotic insulator, a magnetic insulator with magneto-electric parameter $\theta=\pi$, which we
label here as axion insulator.}
\label{FigPhaseDiagram}
\end{figure}

We also mention the possibility of an exotic insulating phase emerging when
the Dirac points annihilate in pairs, as the correlation are reduced. This
phase shows a topological magnetoelectric effect\cite%
{Zhang}, captured by the
magneto-electric parameter $\theta=\pi$, whose value is protected by the
inversion symmetry. Since it is analogous to the axion vacuum in particle
physics\cite{Wilczek}, so we call it the $\theta=\pi$ Axion insulator. Although our LSDA+U+SO calculations find that a metallic phase
intervenes before this possibility is realized, we note that LDA
systematically underestimates gaps, so this scenario could well occur in
reality. Finally, we mention that modest magnetic fields could induce a
reorientation of the magnetic moments, leading to a metallic phase. Our
results are summarized in the phase diagram Figure \ref{FigPhaseDiagram}.
Previous studies considered ferromagnetism \cite{LSDA+U Y2Ir2O7}, and
structural distortion \cite{Kim} in iridates.

Our calculations suggest that new functionalities such as controlling
electrical properties via magnetic textures, and field induced metallic
states can be realized in these materials, with implications for
spintronics, magneto-electrical and magneto-optical devices.

\section{Method}

We perform our electronic structure calculations based on local spin density
approximation (LSDA) to density functional theory (DFT) with the
full--potential, all--electron, linear--muffin--tin--orbital (LMTO) method%
\cite{FP-LMTO}. We use LSDA+U scheme\cite{LDA+U} to take into account the
electron--electron interaction between Ir 5\textit{d} electrons and vary
parameter U between 0 and 3 eV for Ir 5d electrons to see what effects the
on site Coulomb repulsion would bring to the electronic structure of
Iridates. In general, we expect that U can be somewhere between 1 and 2 eV
for the extended 5d states. When the \textit{A} site is a rare earth
element, we also add the Coulomb interaction for the localized 4\textit{f}
electrons and use U = 6 eV. We use a 24$\times $24$\times $24 k--mesh to
perform Brillouin zone integration, and switch off symmetry operations in
order to minimize possible numerical errors in studies of various
(non--)collinear configurations. We use experimental lattice parameters\cite%
{Taira 2001 Ir-227} in all set ups.

Throughout, we exploit inversion symmetry which constrains the phase
diagram, by tracking wavefunction parities at time reversal invariant
momenta. Near electronic phase transitions, a low energy $k.p$ theory is
developed to understand qualitative features of the neighboring phases.
Finally, topological band theory based on momentum space Berry connections
is utilized in deducing the physical properties of the phases.

\section{Magnetic Configuration}

We first study magnetic configuration and discuss our results for Y$_{2}$Ir$%
_{2}$O$_{7}.$Since the strength of the spin orbit (SO) coupling is large for
Ir 5\textit{d} electrons, and leads to insulating behavior in Sr$_{2}$IrO$%
_{4}$\cite{Sr2IrO4-1}, we perform the LSDA+U+SO calculations. There are four
Ir atoms inside the unit cell forming a tetrahedral network as shown in
Fig.1which is geometrically frustrated. Thus, we carry out several
calculations with the initial state to be (i) ferromagnetic, with moment
along (100), (111), (110) or (120) directions (ii) antiferromagnetic with
two sites in a tetrahedron along and other two pointed oppositely to the
directions above; non-colinear structures (iii) "all--in/out" pattern (where
all moments point to or away from the centers of the tetrahedron see Fig 1),
(iv) "2--in/2--out" (two moments in a tetrahedron point to the center of
this tetrahedron, while the other two moments point away from the center,
i.e. the spin--ice\cite{Spin ice Ln2Ti2O7} configuration), and (v)
"3--in/1--out" magnetic structures.

We find that the "all--in/out" configuration is the ground state. Different
from other magnetic configurations, during the self--consistency the
"all--in/out" state will retain their initial input direction; thus, there
is no net magnetic moment. This is consistent with the absence of the
magnetic hysteresis in experiments\cite{Taira 2001 Ir-227}. The
all-in/all-out magnetic configuration was predicted to occur in pyrochlore
antiferromagnets with Dzyaloshinsky-Moriya (D-M) interactions\cite%
{DMPyrochlore}. Symmetry dictates the form of D-M interactions except for
the sign, which leads to two cases, direct and indirect D-M. The
all-in/all-out state is the unique ground state for the former while the
indirect D-M ground state is a coplanar state with the four spins being
either antiparallel of orthogonal\cite{DMPyrochlore}. We find that the
indirect D-M pattern has higher energy than the all-in/all-out state. This
provides further evidence that the all-in/all-out spin configuration is the
natural ground state.

The next lowest energy configuration is the ferromagnetic state.
Interestingly, the rotation of magnetization does not cost much energy
despite strong SO interactions. The (111) direction is found to be lowest
ferromagnetic state, but the energy difference between this and the highest
energy (001)\ state is just about 4.17 meV per unit cell. Also, all of them
produce a considerable net magnetic moment in contrast to the experiment
\cite{exp 2001 Ir-227, Taira 2001 Ir-227, Ir-227}. Our findings are
summarized in Table I for a typical value of U=1.5 eV, and similar results
are found for other values of U in the range from 0 to 3 eV. We find that
the energy difference between the ground and several selected excited states
with different orientations of moments is small. Therefore, modest magnetic
fields may induce a transition into the ferromagnetic state.

\begin{figure}[tbp]
\includegraphics[width=0.3\textwidth]{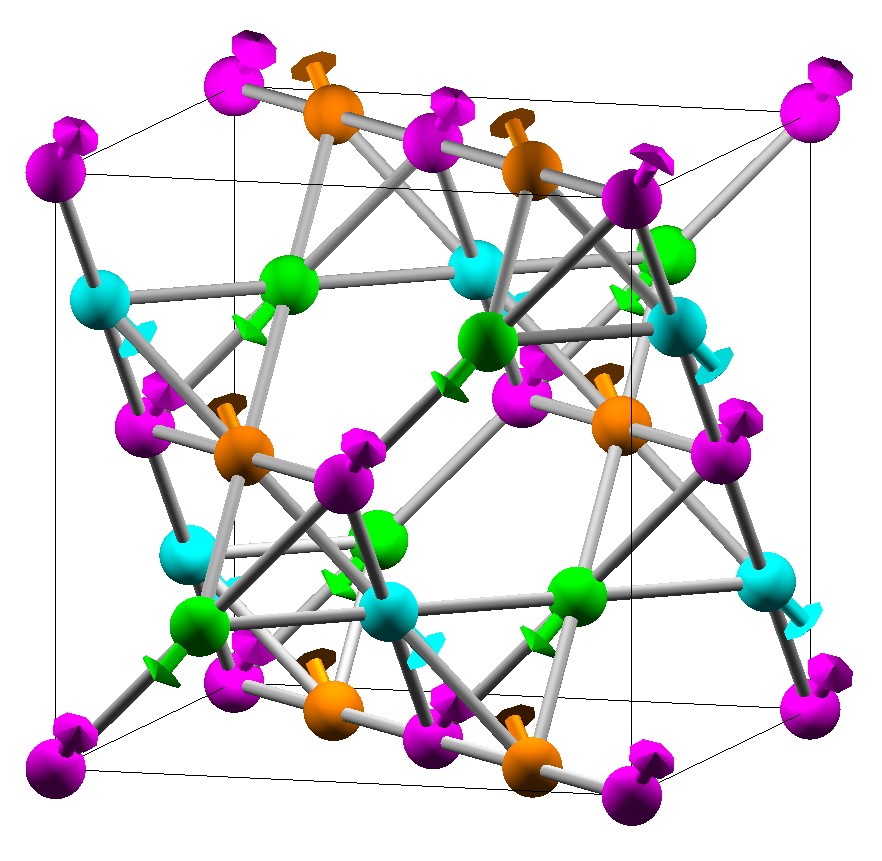}
\caption{The pyrochlore crystal structure showing Ir corner sharing tetrahedral network and
the magnetic configuration corresponding to the "all--in/all--out" alignment of
moments. A degenerate state is obtained on reversing the moments.}
\label{fig:all-in}
\end{figure}

.

\begin{table}[tbp]
\caption{The spin $\langle S\rangle $ and orbital $\langle O\rangle $\
moment (in $\protect\mu _{B}$), the total energy E$_{tot}$ (in meV) for several selected magnetic
configurations of Y$_{2}$Ir$_{2}$O$_{7}$ as calculated using LSDA+U+SO\
method with U=1.5 eV. The IDM is a coplanar configuration predicted for one sign of D-M interactions in Ref. \cite{DMPyrochlore}}%
\begin{tabular}{cccccc}
\hline
Configuration: & (001) & (111) & 2--in/2--out & IDM & all--in/out \\ \hline
$\langle $S$\rangle $ & 0.08 & 0.10 & 0.09 & 0.06 & 0.13 \\
$\langle $O$\rangle $ & 0.09 & 0.10 & 0.07 & 0.06 & 0.12 \\
E$_{tot}$(meV) & 5.47 & 1.30 & 3.02 & 2.90 & 0.00  \\ \hline
\end{tabular}%
\end{table}

\section{Electronic Phases and Wavefunction Parities}

We now discuss electronic properties of Iridates that emerge from our
LSDA+U+SO calculations. A variety of phases ranging from normal metal at
small $U$ to Dirac semi-metallic at intermediate $U\sim 1.5$ eV and Mott
insulating phase at $U$ above 2 eV with non--collinear magnetic "all-in/out"
ordering are predicted. Since pressure or chemical substitution may alter
the screening and the electronic bandwidth resulting in changes in U we
expect that these phases can be observed experimentally in iridates.

The basic features of the electronic structure can be understood by
considering each of four Ir atoms in pyrochlore lattice which is
octahedrally coordinated by six O\ atoms. This makes the Ir 5\textit{d}
state split into doubly degenerate e$_{g}$ and triply degenerate t$_{2g}$
states. Due to the extended nature of Ir 5\textit{d} orbital, the
crystal--field splitting between t$_{2g}$ and e$_{g}$ is large with the e$%
_{g}$ band to be 2 eV higher than the Fermi level. The bands near the Fermi
level are mainly contributed by Ir t$_{2g}$ with some mixing with O 2\textit{%
p} states. SO coupling has a considerable effect on these t$_{2g}$ states:
it lifts their degeneracy and produces quadruplet with $J_{eff}=3/2$ and
doublet with $J_{eff}=1/2$ with the latter lying higher in energy.
Alternatively, one can view the d--shell in strong SO limit which splits the
10 fold degenerate state onto lower lying $J=3/2$ and higher lying $J=5/2$
multiplets. A subsequent application of the cubic crystal field would leave
the $J=3/2$ multiplet degenerate but would split the 6--fold degenerate $%
J=5/2$ state onto $\Gamma _{7}$ doublet and $\Gamma _{8}$ quadruplets. Since
Ir occurs in its 4+ valence, its 5 electrons would fill completely $J=3/2$
subshell and put an additional one electron into the $\Gamma _{7}$ doublet
thought as the state with $J_{eff}=1/2$ in some previous work \cite%
{Sr2IrO4-1}. For the solid, we thus expect 8 narrow energy bands at
half--filling to appear in the vicinity of the chemical potential that
correspond to the $\Gamma _{7}$ doublets of four Ir atoms.

\begin{figure}[tbp]
\includegraphics[width=0.5\textwidth]{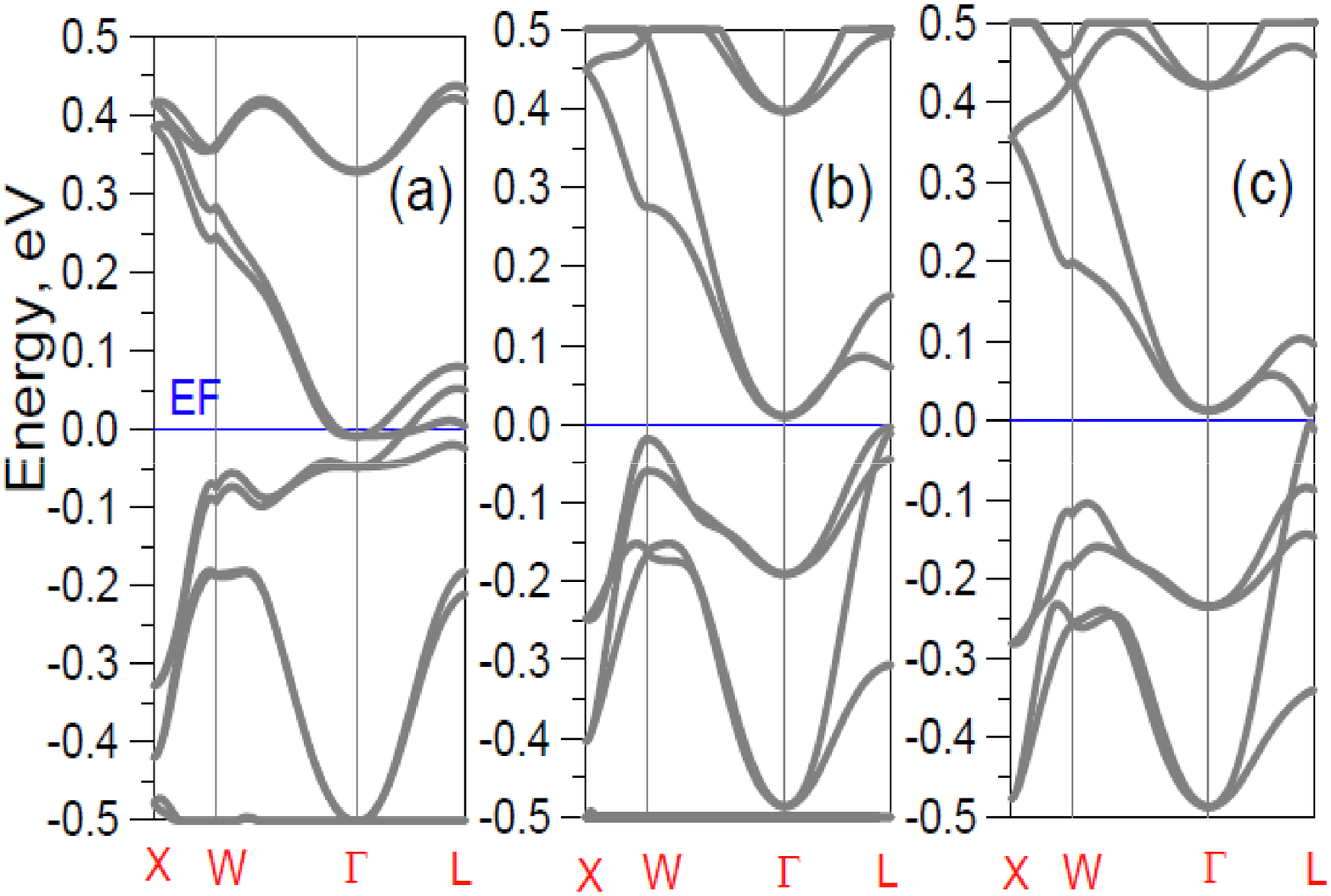}
\caption{Evolution of electronic band structure of Y2Ir2O7 shown along high symmetry directions, calculated using
LSDA+U+SO method with three different values of U equal (a) 0, (metallic) (b) 1.5 eV,
and (c) 2 eV. (Insulator with small gap). The Dirac point that is present in case (b), is not visible along high symmetry lines.}
\label{fig:Bands}
\end{figure}

The precise behavior of these electronic states depends on magnetic
configuration. Our band structure calculations for collinear alignments of
moments show metallic bands regardless the value of $U$ that we use in our
simulations.
On the other hand, we find that the electronic states for the non--collinear
"all-in/out" magnetic state depend strongly on the actual value of $U$ used in
the calculation. In particular, we predict that when $U$ is less than 1 eV,
the ground state is a normal metal while if $U$ is about 1.8 eV or larger, we
find the band structure to be insulating with an energy gap whose value
depends on $U$.
\begin{figure}[h!]
\begin{center}
\includegraphics[scale=0.125]{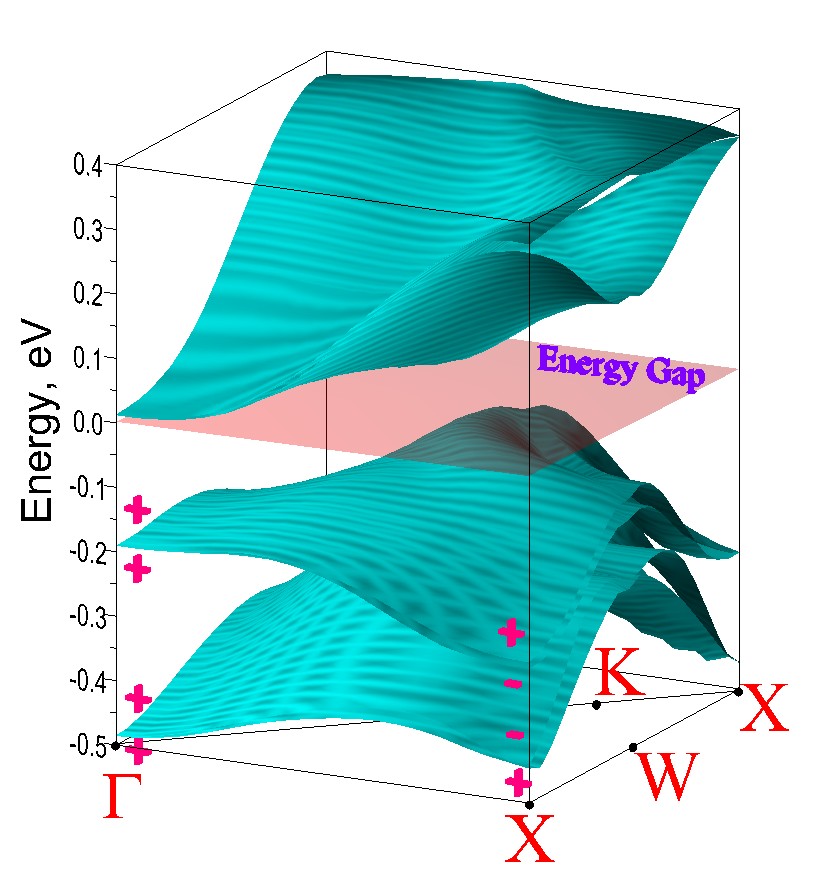}%
\includegraphics[scale=0.125]{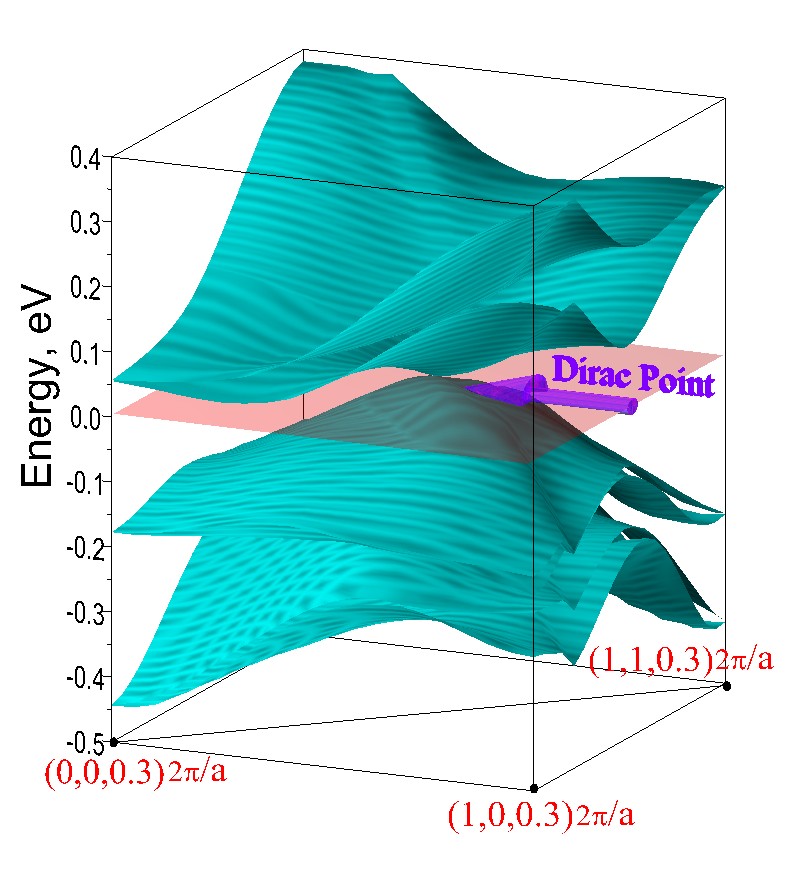}
\end{center}
\par
\label{fig:Dirac}
\caption{Calculated energy bands in the vicinity of the Fermi level using
LSDA+U+SO method with U=1.5 eV. on left: corresponds to plane $k_z=0$ with band
parities shown; (b) corresponds to plane $k_z=0.3 2\protect\pi/a$ where
Dirac point is predicted to exist. The shaded plane is at the Fermi level.}
\end{figure}


{\underline {\emph{Weak Correlations:}}} An interesting recent study
proposed a tight--binding model for the non-magnetic phase of the iridates,
which was a topological insulator \cite{Leon}, a natural phase on the
pyrochlore lattice \cite{Leon, Franz}. Our LDA\ studies of the realistic
electronic structures contradict this, instead we find a metallic phase (see Fig. \ref{fig:Bands}a). One
can understand the discrepancy by analyzing the structure of energy levels
at the $\Gamma$ point (Brillouin Zone center) for the low energy 8--band
complex, composed of the four $J_{eff}=1/2$ states. In \cite{Leon}, these
appear with degeneracies 4,2,2 which after filling with 4 electrons results
in an insulating band structure. Our study of the non--magnetic state using
LDA+SO method (with no U) results, on the other hand, in the sequence 2,4,2
of degeneracies, which is necessarily metallic assuming 4 levels are filled.
Thus, one needs either magnetic order, or a structural transition\cite{Kim},
to recover an insulating phase.

\underline{\emph{Strong Correlations and the Mott Limit:}} When $U>1.8$eV, an insulating
band structure is obtained with the all-in/all-out magnetic configuration,
as shown in the figure \ref{fig:Bands}c. Indeed, it remains qualitatively
similar on increasing $U$ to large values, where a site localized moment is
expected, i.e. the Mott insulator. This can be further verified by
calculating the parity eigenvalues. Note that all the magnetic structures
considered above preserve \emph{inversion} (or parity) symmetry. Thus, if we
pick e.g. an Iridium atom as the origin of our coordinate system, then
inversion $\mathbf{r}\rightarrow -\mathbf{r}$ leaves the crystal structure
and magnetic pattern invariant. This implies a relation between crystal
momenta $\pm \mathbf{k}$. At special momenta, called TRIMs (Time Reversal
Invariant Momenta), that are invariant under inversion, we can label states
by parity eigenvalues $\xi =\pm 1$. In the Brillouin zone of the FCC lattice
these correspond to the $\Gamma =(0,\,0,\,0)$, and $X,\,Y,\,Z$ [=$2\pi
(1,\,0,\,0)$ and permutations] and four $L$ points [$\pi (1,\,1,\,1)$ and
equivalent points]. These parities are very useful to study the evolution of
the band structure. The TRIM parities of the top four occupied bands, in
order of increasing energy, are shown in Table 2. Note, although by symmetry
all $L$ points are equivalent, the choice of inversion center at an Iridium
site singles out one of them, $L^{\prime }$. With that choice the parities
at the two sets of $L$ points are the opposite of one another. The parities
of the all-in/out state remains unchanged above $U>U_{c}\sim 1.8$ eV, and is
shown in the top row under $U=2$ eV. It is readily seen that these parities
arise also for a site localized picture of this phase, where each site has
an electron with a fixed moment along the ordering direction. Due to the
possibility of such a local description of this magnetic insulator, we term
it the Mott phase.

\begin{table}[tbp]
\caption{Calculated parities of states at Time Reversal Invariant Momenta (TRIMs) for several
electronic phases of the iridates. Only the top four filled levels are shown in order of increasing energy.}%
\label{table:Parities}
\begin{tabular}{|l||c|c|c|c|}
\hline
Phase & $\Gamma $ & $X,\,Y,\,Z$ & $L^{\prime }$ & $L\,(\times 3)$ \\ \hline
U=2.0, all-in (Mott) & + + + + & + - - + & + - - - & - + + + \\
U=1.5, all-in (Dirac) & + + + + & + - - + & + - - + & - + + - \\
U=2.0, 111-Ferro & + + + + & + - - + & + - + - & - + - + \\ \hline
\end{tabular}%
\end{table}
\bigskip

\begin{figure}[tbp]
\includegraphics[width=0.3\textwidth]{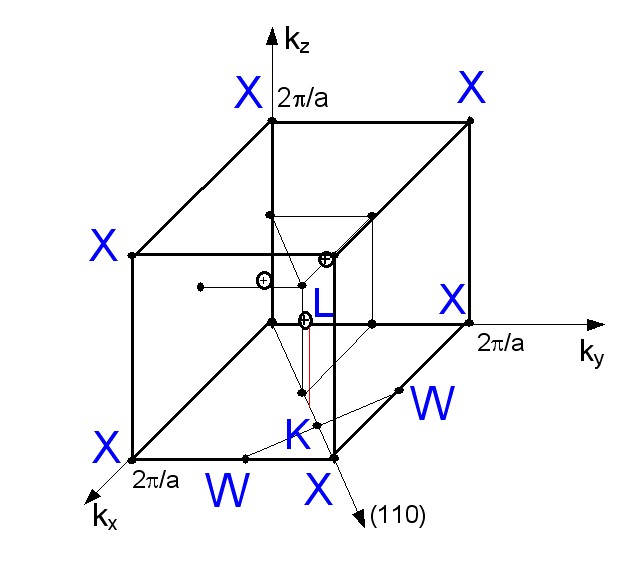}
\caption{Brillouin zone of the FCC lattice with locations of Dirac points
(shown by + signs denoting their "positive" chiral charges) as found by our
LSDA+U+SO calculation with U=1.5 eV for Y2Ir2O7.}
\label{fig:DiracPointLocation}
\end{figure}

\underline{\emph{Intermediate Correlations:}} For the same all-in/out magnetic
configuration, at smaller $U=1.5$ eV, the band structure along high symmetry
lines also appears insulating \ref{fig:Bands}b, and at first sight one may
conclude that this is an extension of the Mott insulator. However, a closer
look at the parities reveals that a phase transition has occurred. A pair of
levels with opposite parity are exchanged at the $L$ points. Near this
crossing point it can readily be argued that only one of the two adjacent
phases can be insulating\cite{SupplementaryMaterial}. Since the large $U$
phase is found to be smoothly connected to a gapped Mott phase, it is
reasonable to assume the smaller $U$ phase is the non--insulating one. This
is also borne out by the LSDA+U+SO band structure. A detailed analysis
perturbing about this transition point (the $k.p$ expansion see \cite%
{SupplementaryMaterial}) allows us to show that a Dirac semi-metal is
expected for intermediate $U$, with $6$ Dirac nodes about every $L$ point.
Indeed, in the LSDA+U+SO band structure, we find a 3 dimensional Dirac
crossing located within the $\Gamma $XL plane of the Brillouin zone. This is
illustrated in Fig.\ref{fig:DiracPointLocation} and corresponds to the
k--vector $(0.52,0.52,0.3)2\pi /a$. There also are two additional Dirac
points in the proximity of the point L related by symmetry. When U
increases, these points move toward each other and annihilate all together
at the L point close to $U=1.8$ eV. This is how the Mott phase is born from
the Dirac phase. Since we expect that for Ir 5d states the actual value of
the Coulomb repulsion should be somewhere within the range 1eV $< U<$ 2eV we
thus conclude that the ground state of the Y$_{2}$Ir$_{2}$O$_{7}$ is most
likely the semi-metallic state with the Fermi surface characterized by a set
of Dirac points but in proximity to a Mott insulating state. Both phases can
be switched to a normal metal if Ir moments are collinearly ordered by a
magnetic field. 

\subsection{Topological Dirac semi-metal}

The effective Hamiltonian in the vicinity of a Dirac momentum $\mathbf{k}=%
\mathbf{k_0}+\mathbf{q}$ is:
\begin{equation}
H_D = \sum_{i=1}^3\mathbf{v}_i\cdot \mathbf{q} \sigma_i  \label{HD}
\end{equation}
, where energy is measured from the chemical potential, and $\sigma_i$ are
the three Pauli matrices. The three velocity vectors $\mathbf{v}_i$ are
generically non-vanishing and linearly independent. The energy dispersion is
$\Delta E=\pm \sqrt{\sum_{i=1}^3(\mathbf{v}_i\cdot \mathbf{q})^2}$. Note, by
inversion symmetry, there must exist Dirac points at both $\mathbf{k_{0}}$
and $-\mathbf{k_{0}}$, whose velocity vectors are reversed. One can assign a
\emph{chirality} (or chiral charge) $c=\pm 1$ to the fermions defined as $c=%
\mathrm{sign}(\mathbf{v}_{1}\cdot \mathbf{v}_{2}\times \mathbf{v}_{3})$, so
Dirac points related by inversion have opposite chirality. Note, since the $%
2\times2$ Pauli matrices appear, our Dirac particles are two component
fermions. In contrast to regular four component Dirac fermions, it is not
possible to introduce a mass gap. The only way to eliminate them is if they
meet with another 2 component Dirac dispersion in the Brillouin Zone, but
with opposite chiral charge. Thus they are topological objects. We note that
near each $L$ point there are three Dirac points related by the three fold
rotation, which have the same chiral charge. Fig.\ref{fig:DiracPointLocation}
denotes those points as "+" dots. Another three Dirac points with opposite
chirality, related by inversion. Thus, there are 24 Dirac points in the
whole Brillouin zone. Since they are all related by symmetry, they are at
the same energy. The chemical potential is fixed to be at the Dirac point
energy as verified in the microscopic calculation. The Fermi velocities at
the Dirac point are found to be typically an order of magnitude smaller than
in graphene. We briefly note that this Dirac semimetal is a critical state
with power law forms for various properties, which will be described in more
detail elsewhere. For example, the density of states $N(E)\propto E^2$. The
small density of states makes it an electrical insulator at zero
temperature. For a single node with isotropic velocity $v$, the a.c.
conductivity in the free particle limit of the clean system is $%
\sigma(\Omega)=\frac{e^2}{12h}\frac{|\Omega|}{v}\tanh{|\Omega|/4k_BT}$.

{\underline {\emph{Surface States:}}} We now discuss surface states that are
associated with the presence of the two component Dirac fermions. We first
note that they behave like 'magnetic' monopoles of the Berry flux whose
charge is given by the chirality.The Berry connection, a vector potential in
momentum space, is defined by ${\mathcal{A}}(\mathbf{k})=\sum_{n=1}^Ni%
\langle u_{n\mathbf{k}}|\nabla_\mathbf{k}|u_{n\mathbf{k}}\rangle$ where $N$
is the number of occupied bands. As usual, an analog of the magnetic field,
the Berry flux, is defined as ${\mathcal{F}}=\nabla_{\mathbf{k}}\times
\mathcal{A}$. Now consider energy eigenstates at the Fermi energy (taken to
be at $E=0$). In the bulk, this corresponds to the set of Dirac points,
hence the bulk Fermi surface is a collection of Fermi points. However, in
the presence of a surface (say the plane $z=0$), new low energy states may
be generated. We show that these will occur along a curve in the surface
Brillouin zone as is illustrated in Fig. 5. The end points of this curve
occur at the bulk Fermi point momenta, projected onto the surface Brillouin
Zone. Also, the curve connects Dirac nodes with opposite monopole charge. If
more than one Dirac node projects to the same surface momentum, the sum of
the monopole charges should be considered. This is argued by showing that
there must be Fermi arcs on the surface Brillouin zone emanating from the
projection $(k_{0x},k_{0y})$ of the monopole as argued below.

\begin{figure}[tbp]
\includegraphics[width=0.5\textwidth]{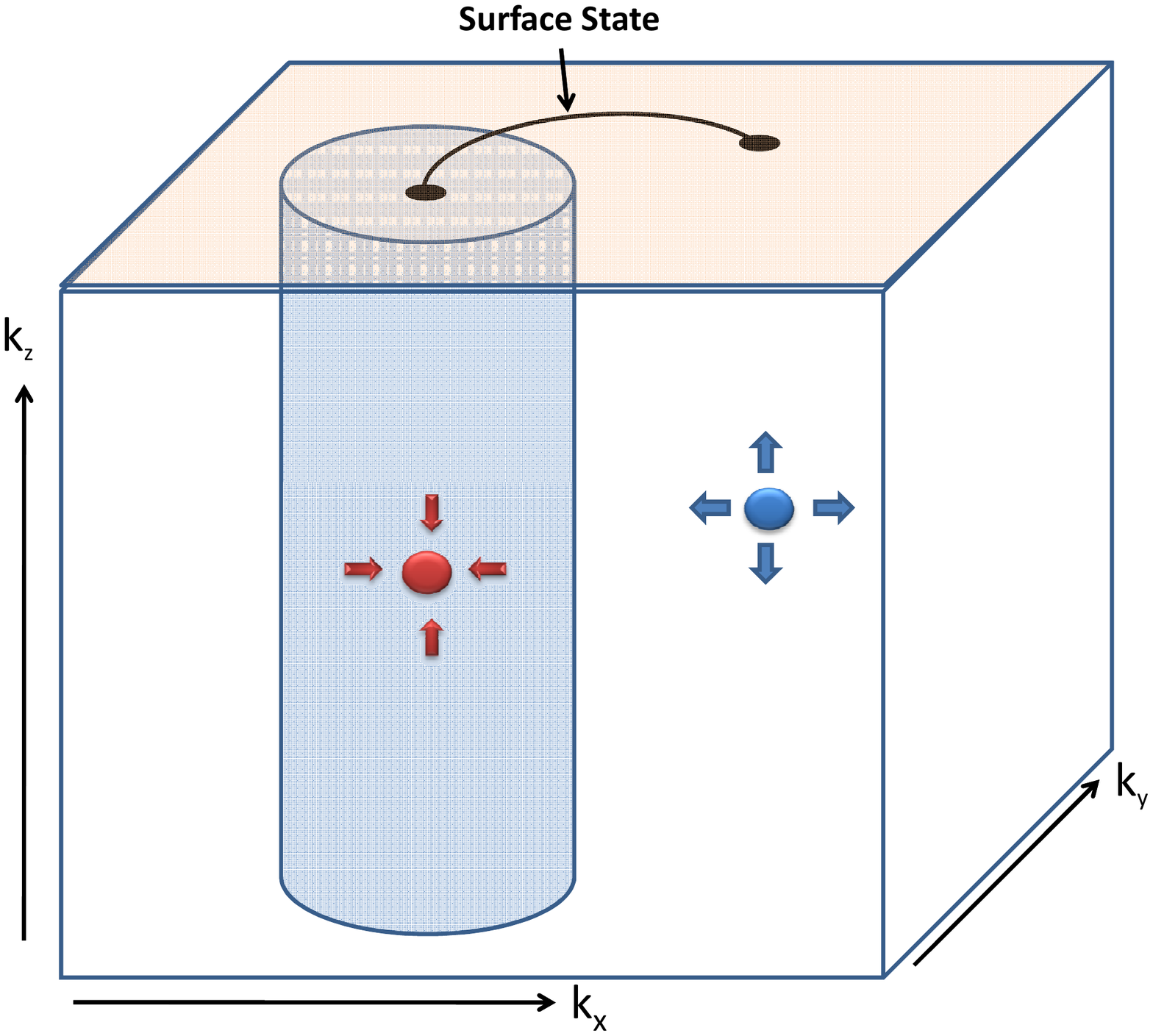}
\caption{Illustration of surface states arising from bulk Dirac points. For simplicity, only a pair of Dirac points with opposite chirality are shown. The imaginary cylinder in momentum space has unit Chern number, due to the Berry monopole at the Dirac point. Hence a surface state must arise, as shown schematically in the same plot. When the Fermi energy is at the Dirac point, a Fermi arc is present which connects the surface momenta of the projected bulk Dirac points
of opposite chirality. }
\label{fig:BZ}
\end{figure}

\emph{Origin of Surface States:} We now prove that the band topology
associated with the Dirac point leads to surface states. Construct a curve
in the surface Brillouin zone encircling the projection of the bulk Dirac
point, which is traversed counterclockwise  as we vary the parameter $%
\lambda :0\rightarrow 2\pi $; $\mathbf{k_{\lambda }}=(k_{x}(\lambda
),k_{y}(\lambda ))$ (see Fig. \ref{fig:BZ}). We show that the energy $%
\epsilon _{\lambda }$ of a surface state at momentum $\mathbf{k_{\lambda}}$
crosses $E=0$. Consider $H(\lambda ,k_{z})=H(\mathbf{k_{\lambda }},k_{z})$,
the gapped Hamiltonian of the two dimensional subsystem defined by this
curve. The two periodic parameters $\lambda ,\,k_{z}$ define the surface of
a torus in momentum space (see Fig. \ref{fig:BZ}). The Chern number of this two dimensional band
structure is given by the Berry curvature integration: $\frac{1}{2\pi }\int {%
\mathcal{F}}dk_{z}d\lambda $ which, by Stokes theorem, simply corresponds to
the net monopole density enclosed within the torus. This is obtained by
summing the chiralities of the enclosed Dirac nodes. Consider the case when
the net chirality is unity, corresponding to a single enclosed Dirac node.
Then, the two dimensional subsystem defines a quantum Hall insulator with
unit Chern number. When defined on the half space $z<0$, this corresponds to
putting the quantum Hall state on a cylinder, and hence we expect a chiral
edge state. Its energy $\epsilon _{\lambda }$ spans the band gap of the
subsystem, as $\lambda $ is varied. Hence, this surface state crosses zero
energy somewhere on the surface Brillouin zone $\mathbf{k_{\lambda _{0}}}$.
Such a state can be obtained for every curve enclosing the Dirac point.
Thus, at zero energy, there is a Fermi line in the surface Brillouin zone,
that terminates at the Dirac point momenta (see Fig. \ref{fig:BZ}). An arc beginning on a Dirac
point of chirality $c$ has to terminate on a Dirac point of the opposite
chirality. Clearly, the net chirality of the Dirac points within the $%
(\lambda ,\,k_{z})$ torus was a key input in determining the number of these
states. If Dirac points of opposite chirality line up along the $k_{z}$
direction, then there is a cancelation and no surface states are expected.

For $U=1.5$ eV, a Dirac node is found to occur at the momentum $%
(0.52,0.52,0.31)2\pi /a$ and equivalent points (see Fig.3). They can be
thought of as occuring on the edges of a cube, with a pair of Dirac nodes of
opposite chirality occupying each edge, as, e.g., the points $%
(0.52,0.52,0.31)2\pi /a$ and $(0.52,0.52,-0.31)2\pi /a.$ For the case of $%
U=1.5$ eV, the sides of this cube have the length $0.52(4\pi /a)$. Thus, the
$(111)$ and $(110)$ surfaces would have surface states connecting the
projected Dirac points. If, on the other hand we consider the surface
orthogonal to the $(001)$ direction, it would lead to the Dirac points of
opposite chirality being projected to the same surface momentum, along the
edges of the cube. Thus, no protected states are expected for this surface.

\begin{figure}[tbp]
\includegraphics[width=0.4\textwidth]{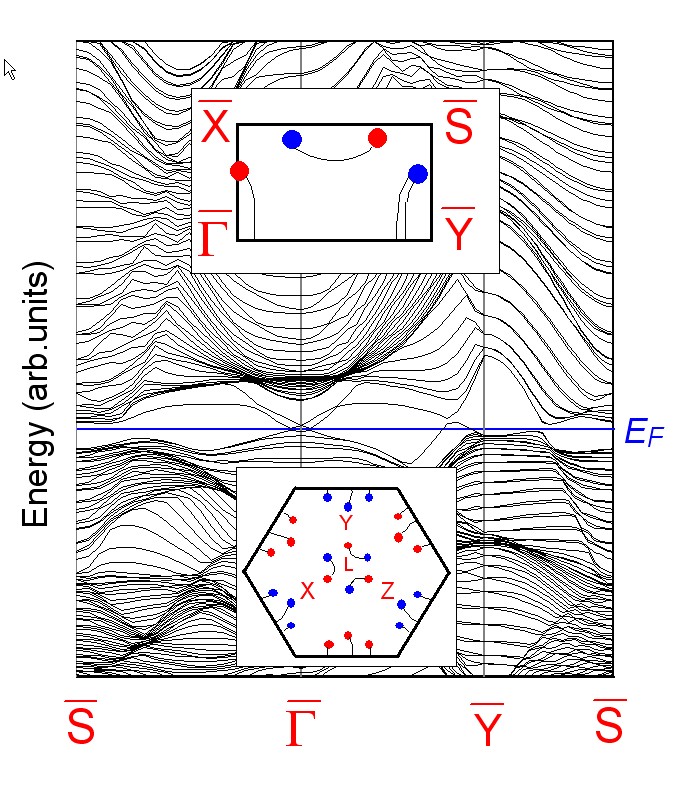}
\caption{Calculated surface energy bands corresponding to (110) surface of
the pyrochlore iridate Y2Ir2O7. A tight binding approximation has been used
to simulate the bulk band structure with 3D Dirac point as found by our
LSDA+U+SO calculation. The plot corresponds to diagonalizing 128 atoms slab
with two surfaces. The top inset shows the deduced Fermi arcs connecting
projected bulk Dirac points of opposite chirality. The bottom inset shows a
sketch how these Fermi arcs are expected to behave for the (111) surface.}
\label{fig:Surfacestates}
\end{figure}


\emph{Model Calculation:} To verify these theoretical considerations, we
have constructed a tight binding model which has features seen in our
electronic structure calculations for YIr$_{2}$O$_{7}$. We consider only t$%
_{2g}$ orbitals of Ir atoms in the global coordinate system. Since Ir atoms
form tetrahedral network (see Fig. \ref{fig:all-in}), each pair of nearest
neighboring atoms forms a corresponding $\sigma -$like bond whose hopping
integral is denoted as $t$ and other two $\pi -$like bonds whose hopping
integrals are denoted as $t^{\prime }$. To simulate the appearance of the
Dirac point it is essential to include next--nearest neighbor interactions
between t$_{2g}$ orbitals which are denoted as $t^{\prime \prime }.$ With
the parameters $t=0.2,t^{\prime }=0.5t,t^{\prime \prime }=-0.2t$, the value
of the on--site spin--orbit coupling equal to $2.5t$ and the applied
on--site splitting between spin up and spin down states equal to 0.1
referred to the local quantization axis which simulates our non--collinear
'all--in/out' configuration we can model both the bulk Dirac metal state and
its surface. The calculated (110)\ surface band structure for the slab of
128 atoms together with the sketch of the obtained Fermi arcs is shown in
Fig. \ref{fig:Surfacestates}. Notice that since the slab calculation
involves two surfaces, the corresponding surface states and Fermi arcs for
both surfaces are generated. We also display the expected surface states for
the (111) surface. Note, no special surface states are expected for the
(001) surface.

\section{Other Topological Phases}

We recall that topological insulators are non-magnetic band insulators with
protected surface states \cite{TI}. Time reversal symmetry is required in
the bulk to define these phases. When the surface states are eliminated by
adding, for example magnetic moments only on the surface, a quantized
magneto-electric response is obtained, where a magnetic field induces a
polarization: $\mathbf{P} = \theta \frac{e^2}{2\pi h} \mathbf{B}$, with the
coefficient $|\theta|$ \cite{Zhang} is only defined modulo $2\pi$. Under
time reversal, $\theta \rightarrow -\theta$. Apart from the trivial solution
$\theta=0$, the ambiguity in the definition of $\theta$ allows also for $%
\theta=\pi$. For topological insulators $\theta=\pi$.


In magnetic insulators, $\theta$ is in general no longer quantized \cite{Ari}%
. However, when inversion symmetry is retained, $\theta$ is quantized again,
since inversion also changes its sign. Thus again $\theta=0,\,\pi\, \mathrm{%
mod }2\pi$, and an insulator with the latter value will be termed a $%
\theta=\pi$ axion insulator.

Which is the appropriate description of the pyrochlore iridate phases we
have described? For the Mott insulator, at large $U$, the charge physics
must be trivial and so we must have $\theta=0$.
Next, since the Dirac semi metal phase is gapless in the bulk, $\theta$ is
ill defined. However, we note that on reducing $U$, the location of the
Dirac points shift, with nodes of opposite chirality approaching each other.
If these meet and annihilate, then one recovers a gapped phase in the low $U$
regime. However, in the process the resulting phase will have $\theta=\pi$. Indeed, the
presence of the intervening Dirac phase can be deduced from the requirement
that $\theta$ has to change between these two quantized values. As described
elsewhere\cite{AriZhangAV}, the condition for $\theta=\pi$ when deduced from
the parities, turns out to be the same as the Fu-Kane formula, for time
reversal symmetric insulators \cite{FuKane,QiZhang}, i.e. if the total
number of filled states of negative parity at all TRIMs taken together is twice an odd
integer, then $\theta=\pi$. Otherwise $\theta=0$. The small $U$ insulator has the same
parities as the Dirac semi-metal, since the Dirac points annihilate away from a TRIM. From Table \ref{table:Parities} we can see that indeed this corresponds to $\theta=\pi$, since there are 14 negative parity filled states, while the Mott insulator corresponds to $\theta=0$, having 12 negative parity filled states.

Unfortunately, within our LSDA+U+SO calculation, a
metallic phase intervenes on lowering $U \leq 1.0$eV, before the Dirac
points annihilate to give the axion insulator. We point out this possibility
nevertheless, since LDA systematically underestimates the stability of such
gapped phases. Moreover it provides an interesting example of a pair of
insulators with the same magnetic order, but which are nevertheless
different phases. Thus, the Mott insulator is distinct from the smaller $U$
`Slater' insulator, unlike in many other cases where they are smoothly
connected to one another. Inversion symmetry is critical in preserving this distinction.

In summary, a theoretical phase diagram for the physical
system is shown in Figure \ref{FigPhaseDiagram} as a function of $U$ and
applied magnetic field, which leads to a metallic state beyond a critical
field. The precise nature of these phase transformations are not addressed in
the present study

\section{Acknowledgements}
A.V. thanks L. Balents, J. Orenstein and R. Ramesh for insightful discussions. X.W. acknowledges support by National Key Project for Basic Research of China (Grant No. 2006CB921802, and 2010CB923404), NSFC under Grant No. 10774067, and 10974082. S.S. acknowledges support by DOE SciDAC Grant No. SE-FC02-06ER25793 and thanks Nanjing University for the kind hospitality during his visit to China. X.W. and S.S also acknowledge support from Kavli Institute for Theoretical Physics where this work has been initiated. The work at Berkeley was supported by the Office of Basic Energy Sciences, Materials Sciences Division of the U.S. Department of Energy under contract No. DE-AC02-05CH1123.
\section{Appendix}
\subsection{I. Effective k.p Theory and Intervening Dirac Metal Phase}

Consider a pair of states at the $L$ point which have opposite parity, and
cross each other as we tune $U$. We want to understand what happens to the
band structure.

The $L$ point has two symmetries which do not change its crystal momentum.
First of course is inversion, and we can label states by the eigenvalues $%
P=\pm 1$. The second is 120$^o$ rotations about a line joining $L-\Gamma$.
There are three possible eigenvalues which we call $s=-1/2,\,1/2,\,3/2$. So,
any state at this point can be labeled by $\{P,\,s\}$. Now consider writing
the effective $2\times 2$ Hamiltonians for a pair of states that are near
each other in energy:

\begin{enumerate}
\item \emph{At the $L$ point:} Since we have inversion symmetry, the two
states taken to be eigenvalues of $\tau_z=\pm1$, cannot mix. Hence the
effective Hamiltonian is
\begin{equation*}
H(L) = \Delta\tau_z
\end{equation*}

where the coefficient $\Delta$ changes sign when the levels pass through
each other. Note, the $s$ quantum number of the two levels is irrelevant
here.

\item \emph{Along the $\Gamma-L$ Direction:} We still have the quantum
number $s$, but not $P$, since that inverts the momentum. Denoting by $k_z$
the momentum along this line deviating from the $L$ point, we have two
cases. If the $s$ quantum number of the two levels is different, they still
cannot mix, so the effective Hamiltonian is $H=(\Delta+k_z^2)\tau_z$. Now,
when $\Delta<0$ there are two solutions $k_z=\pm \sqrt{-\Delta}$, where
there are nodes along this $\Gamma-L$ line. You can see this for the $s=1/2$
and $s=3/2$ crossings in the data.

However, if the states have the \emph{same} $s$ quantum number they can mix,
once you move away from $L$. Now the effective Hamiltonian is:
\begin{equation*}
H(\Gamma-L) = (\Delta+k_z^2)\tau_z + k_z \tau_x
\end{equation*}
where the second term arises since inversion is broken on moving away from $%
L $ allowing for mixing. Now, the spectrum is $E=\sqrt{(k_z^2-|%
\Delta|)^2+k_z^2}$, for $\Delta<0$, so despite a level crossing there is no
node along the $\Gamma-L$ line.

\item \emph{General Point in BZ:} In the latter case, does this mean there are no Dirac points?
No - we just need to move off the $\Gamma - L$ line. Let the deviation be $%
\mathbf{k}_\perp$, a 2 vector. The fact the 2 levels have opposite parity
means we need an odd function of $k_\perp$ to induce a matrix element between the levels. And also,
since $\mathbf{k}_\perp$ is a 2 vector, it transforms under the rotation -
the rotationally symmetric form allowed is $\Delta H = k_\perp^3
\cos3\theta\tau_x + k_\perp^3\sin3\theta \tau_y$. Putting this all together
we have the effective Dirac Hamiltonian near the $L$ point:

\begin{equation*}
H(\mathbf{k}) = (\Delta+k_z^2)\tau_z + (k_z +k_\perp^3 \cos3\theta) \tau_x+
k_\perp^3\sin3\theta \tau_y
\end{equation*}
Note, this has the form $A(k)\tau_z+B(k)\tau_x+C(k)\tau_y$. For a node, $%
A=B=C=0$. This occurs if:$A=0$ so $k_z = +\sqrt{-|m|}$, and $C=0$ implies $%
\theta= p\pi/3$, where $p$ is an integer; and finally $k_\perp^3 = k_z$,
when $p=1,3,5$. Similarly for $k_z<0$, the nodes are inverted. In all we
have 6 nodes for this $L$ point, 24 Nodes in all.
\end{enumerate}

Actually this is not the complete expansion. Strictly we should write $A(k)
= m+k_z^2+\alpha k_\perp^2$. This is an effective mass Hamiltonian near the
L point. If $\alpha>0$, then it turns out the conclusions are the same as
above, without this additional term (i.e., there is a transition from
$\theta=\pi$ magnetic axion insulator to Dirac metal on increasing U). \emph{However}, if $%
\alpha<0$, this completely changes the conclusions, as discussed below. In
fact, this turns out to be the physically relevant case according to the
electronic structure calculations for Y$_2$Ir$_2$O$_7$.

Let us assume $\alpha=-1/m_2<0$. Then the effective Hamiltonian:
\begin{equation}
H(\mathbf{k}) = (\Delta+\frac{k_z^2}{2m_1}-\frac{k_\perp^2}{2m_2})\tau_z +
(\beta k_z +k_\perp^3 \cos3\theta) \tau_x+ k_\perp^3\sin3\theta \tau_y
\end{equation}
where a few parameters have been labeled. The Dirac nodes then are at: (i) $%
C=0 \rightarrow \theta=p\pi/3$ and (ii) using this, $B=0 \rightarrow k_z =
\pm k_\perp^3/\beta$ depending on whether we look at $p=$odd or $p=$ even.
Finally, using these relations we have for $A=0$ equation:
\begin{equation*}
\Delta + \frac{k_\perp^6}{2\beta^3m_1}-\frac{k_\perp^2}{2m_2}=0
\end{equation*}

for small $\Delta$ this has the solution $k_\perp^2=2m_2\Delta$. Note, this
has a solution only for $\Delta>0$, i.e. before the gap gets inverted at the
$L$ point on increasing $U$. Thus, in this scenario, there is a Dirac point
only in the small $U$ phase. These Dirac points live along $\theta=p\pi/3$
plane, which is the $k_x=k_y$ plane that contains the points $\Gamma-L-K$
and rotations thereof.



\subsection{II. Summary of Experiments}



We now summarize the experimental facts about the pyrochlore iridates $A_2$Ir%
$_2$O$_7$. Early work revealed that increasing the $A$ ion radius triggered
a metal-insulator transition in the ground state. Thus while $A=$Pr is
metallic, $A=$Y is insulating at low temperatures \cite{exp 2001 Ir-227}.
Reducing the ionic radius is believed to narrow the bandwidth and increase
correlations. Subsequent improvement in sample quality revealed that $A=$Eu,
Sm, Nd also displayed low temperature insulating states\cite{Matsuhira
Ir-227}. In these systems, a metal insulator transition is clearly observed
on cooling (eg. at $T_{MI}=120K$ for $A=$Eu). At the same temperature, a
signature in magnetic susceptibility is also observed, indicative of a
magnetic transition \cite{Taira 2001 Ir-227}. This magnetic signature,
wherein the field cooled and zero field cooled magnetic susceptibilities
separate below the transition temperature, is reminiscent of a spin glass
state. Since this signature is see in $A=$Y, Lu \cite{Taira 2001 Ir-227},
with nonmagnetic $A$ site atoms, it is associated with Ir site moments. In Y$%
_2$Ir$_2$O$_7$, no sharp resistivity signature has been reported at the
magnetic transition, but the resistivity climbs steeply on cooling below
this temperature. Moreover, light hole doping suppresses both the insulating
state and the magnetic transition\cite{Ir-227}. Finally, we note that a
thermodynamic signature of the magnetic transition, a bump in the specific
heat, is observed in clean samples of $A=$Sm \cite{Matsuhira Ir-227}. X-ray
scattering did not observed any structural change below the ordering
transition \cite{Matsuhira Ir-227}, although the presence of new lines in
the Raman spectroscopy \cite{RamanIridate} has been attributed to lowering
of cubic symmetry in $A=$ Eu, Sm but not in $A=$Nd.

We now discuss how the present theoretical description sits with these
facts. We propose that the low temperature state of Y$_2$Ir$_2$O$_7$ (and
also possibly of $A=$ Eu, Sm and Nd iridates) is the Dirac semi-metal, with
all-in/all-out magnetic order. This is broadly consistent with the
interconnection between insulating behavior and magnetism observed
experimentally. It is also consistent with being proximate to a metallic
phase on lowering the correlation strength, such as $A=$Pr. In the clean
limit, a three dimensional Dirac semi-metal is an electrical insulator, and
can potentially account for the observed electrical resistivity. The
noncolinear magnetic order proposed has Ising symmetry and could undergo a
continuous ordering transition. Since this configuration is not frustrated,
it is not clear how spin glass behavior would arise, but the observed
magnetic signature could perhaps also arise from defects like magnetic
domain walls. A direct probe of magnetism is currently lacking and would
shed light on this key question.


\end{document}